\documentclass[journal]{IEEEtran}
\usepackage[utf8x]{inputenc}
\usepackage{paralist}
\usepackage{enumitem}
\usepackage{lipsum}
\usepackage{framed}

\usepackage{color}
\usepackage{bm} 

\usepackage{amsmath}
\usepackage{amsfonts}
\usepackage{amssymb}

\usepackage{stfloats}
\usepackage{graphicx}
\usepackage{epstopdf}
\usepackage[caption=false]{subfig}
\graphicspath{{figures/}}

\makeatother
\makeatletter
\renewcommand{\maketag@@@}[1]{\hbox{\m@th\normalsize\normalfont#1}}%
\makeatother
\newcounter{TempEqCnt}

\usepackage{url}
\usepackage{soul}

\DeclareRobustCommand{\mhl}[1]{%
	\ifmmode\text{\rvp{$#1$}}\else\revisedpart{#1}\fi
}

\newcommand{\rvp}[1]{{\color{blue}{#1}}}

\usepackage{amsthm}



\usepackage{booktabs}
\usepackage{multicol}
\usepackage{threeparttable}
\usepackage{multirow}
\usepackage{tabularx}
\newcolumntype{L}[1]{>{\raggedright\arraybackslash}p{#1}}
\newcolumntype{C}[1]{>{\centering\arraybackslash}p{#1}}
\newcolumntype{R}[1]{>{\raggedleft\arraybackslash}p{#1}}

\usepackage[level]{datetime}  
\newdateformat{ukdate}{\ordinaldate{\THEDAY}~\monthname[\THEMONTH]~\THEYEAR}

\usepackage{algorithm}
\usepackage{algorithmic}
\usepackage{setspace}

\usepackage{hyperref}
\usepackage[nameinlink,noabbrev]{cleveref}

\crefname{equation}{}{}
\crefname{figure}{Fig.}{Fig.}
\crefname{table}{Table}{Table}
\crefname{lemma}{Lemma}{Lemma}
\crefname{prop}{Proposition}{Proposition}
\crefname{thm}{Theorem}{Theorem}
\crefname{defn}{Definition}{Definition}
\crefname{rem}{Remark}{Remark}
\usepackage[square, comma, sort&compress, numbers]{natbib}

\setstcolor{red}

\begin{document}
	
	\title{Network-Assisted Full-Duplex Cell-Free mmWave Massive MIMO Systems with DAC Quantization and Fronthaul Compression}
	
	\author{~Jiamin~Li,~\IEEEmembership{Member,~IEEE,}~Qingrui~Fan,~\IEEEmembership{Student~Member,~IEEE,}~Yu~Zhang,~Pengcheng~Zhu,~\IEEEmembership{Member,~IEEE,}\\Dongming~Wang,~\IEEEmembership{Member,~IEEE,}~Hao Wu,~\IEEEmembership{Member,~IEEE,}~and~Xiaohu~You,~\IEEEmembership{Fellow,~IEEE}

    \thanks{\emph{Corresponding author: Yu Zhang (e-mail: zhangyu@nuist.edu.cn)}}
	\thanks{J. Li, Q. Fan, P. Zhu, D. Wang and X. You are with National Mobile Communications Research Laboratory, Southeast University, Nanjing 210096, China (e-mail: jiaminli, fanqingrui, p.zhu, wangdm, xhyu@seu.edu.cn).}
	\thanks{Yu Zhang is with the Department of Electronic \& Information Engineering, Nanjing University of Information Science \& Technology, Nanjing 210044, China (email: zhangyu@nuist.edu.cn).}
	\thanks{Hao. Wu is with State Key Laboratory of Mobile Network and Mobile Multimedia Technology, ZTE Corporation, Shenzhen, 518055, China (email: wu.hao89@zte.com.cn).}
	}

	\maketitle
	
	\begin{abstract}
		In this paper, we investigate network-assisted full-duplex (NAFD) cell-free millimeter-wave (mmWave) massive multiple-input multiple-output (MIMO) systems with digital-to-analog converter (DAC) quantization and fronthaul compression. We propose to maximize the weighted uplink and downlink sum rate by jointly optimizing the power allocation of both the transmitting remote antenna units (T-RAUs) and uplink users and the variances of the downlink and uplink fronthaul compression noises.  To deal with this challenging problem, we further apply a successive convex approximation (SCA) method to handle the non-convex bidirectional limited-capacity fronthaul constraints. The simulation results verify the convergence of the proposed SCA-based algorithm and analyze the impact of fronthaul capacity and DAC quantization on the spectral efficiency of the NAFD cell-free mmWave massive MIMO systems. Moreover, some insightful conclusions are obtained through the comparisons of spectral efficiency, which shows that NAFD achieves better performance gains than co-time co-frequency full-duplex cloud radio access network (CCFD C-RAN) in the cases of practical limited-resolution DACs. Specifically, their performance gaps with $8$-bit DAC quantization are larger than that with $1$-bit DAC quantization, {\color{blue}which attains a 5.5-fold improvement.}
	\end{abstract}
	

	%
	\IEEEpeerreviewmaketitle

	\section{Introduction}

	\IEEEPARstart{C}ell-free massive multiple-input multiple-output (MIMO) has been widely regarded as a key physical layer technology for the fifth-generation (5G)  and beyond 5G \cite{ngo2017cell,nguyen2020spectral,zhu2021optimization}. {\color{blue}In this promising network architecture, a very large number of remote antenna units (RAUs) are distributed in a wide area to jointly serve a much smaller number of single-antenna user equipments (UEs), which makes cell-free massive MIMO a distributed form of classical centralized massive MIMO \cite{ngo2017cell}}, {\color{red}\cite{buzzi2017cell}}. By implementing simple signal processing at the transmitting and receiving ends, cell-free massive MIMO inherits the advantages of centralized massive MIMO such as channel hardening and beneficial propagation effects \cite{2020Energy}. {\color{red}Although the propagation effects of millimeter-wave (mmWave) communication are limited by rain attenuation, poor foliage penetration, etc., there are still available propagation gain advantages, such as (i) favorable propagation is suitable for short-range ($\le20$ km) communications; (ii) a high degree of frequency reuse is allowed since limited range communications; (iii) mmWave frequency can provide relatively safe communication in the absorption resonance band \cite{marcus2005millimeter}.} 
	
	With a continuous improvement of communication rate requirements, many attempts have been made to exploit less-congested wide bandwidths in the mmWave spectrum \cite{gao2016energy,zhang2020channel,zhang2020hybrid,zhan2021interference}.
	Therefore, an integration of mmWave communications and cell-free massive MIMO is a promising direction for the next generation wireless networks~\cite{femenias2019cell}.
	In particular, the macro-diversity gain achieved by deploying lots of distributed APs compensate for the spotty coverage of mmWave signals.
	To substantially reduce the costs of deployment and power consumption of APs in the cell-free network, low-complexity hybrid analog-digital beamforming designs have been ultilized~\cite{jin2019channel}.

In recent years, co-time co-frequency full-duplex (CCFD) has attracted widespread attention. Through utilizing advanced self-interference cancellation technology, the full-duplex wireless transceiver can transmit and receive the signals simultaneously at the same time-frequency resource block, thereby enabling the double spectral efficiency of the wireless networks \cite{sabharwal2014band,nguyen2020spectral,shu2017high}. However, CCFD is usually limited to small-scale point-to-point communications. When CCFD comes to ultra-dense networks (UDNs), the  cross-link interference (CLI) including  the uplink-to-downlink interference, inter-user interference (IUI) and inter-RAU interference (IRI), severely decrease the performance of the CCFD system based on stochastic geometry analysis \cite{alammouri2016band}, \cite{bi2018fractional}. Therefore, eliminating CLI is an urgent issue for large-scale wireless networks.
	
   To this end, an network-assisted full-duplex (NAFD) cell-free massive MIMO system was originally proposed by Wang \emph{et al} to reduce CLI in the networks through unifying flexible duplex, hybrid-duplex, full-duplex and other duplex schemes~\cite{wang2019performance}. In such a system, {\color{blue}RAUs} exist in pairs that include one transmitting RAU (T-RAU) for downlink communication and one receiving RAU (R-RAU) for uplink communication~\cite{li2020network}. Therein, each RAU is connected to a central processing unit (CPU) where baseband processing is carried out intensively via fronthaul links. Due to the geographically distributed layout of RAUs, NAFD is proven to achieve better spatial diversity than CCFD~\cite{lv2018spectral,xia2021joint,yang2021joint}.

   Motivated by the above discussions, an NAFD cell-free mmWave massive MIMO system is promised to achieve the following advantages: (\romannumeral1) realize the network with truly flexible duplex mode~\cite{zhu2021optimization}; (\romannumeral2) facilitate the bidirectional communication links resource efficiently with CLI cancellation \cite{wang2019performance}; (\romannumeral3) compensate for the orders-of-magnitude increase in free-space pathloss due to the use of higher frequencies with appropriate hybrid analog-digital beamforming \cite{gao2016energy,femenias2019cell}. Consequently, theoretical performance is worth analyzing to demonstrate the superiority of NAFD cell-free mmWave massive MIMO systems.

	
	NAFD cell-free mmWave massive MIMO system has the potential to enhance wireless transmission but it has several limitations. For example, along with an increase of radio frequency (RF) chains with high-resolution digital-to-analog converters (DACs), the power consumption of the proposed NAFD system becomes impractical and costly. Thus, it is more appropriate to adopt a hybrid analog-digital architecture with low-resolution DACs for each RAU so as to promote green communication with low-energy consumption \cite{ribeiro2018energy}. Another bottleneck of NAFD cell-free mmWave massive MIMO system lies in the constraint on the fronthaul compression capacity~\cite{gao2015mmwave}. Unlike the infinite fronthaul capacity in theory, the fronthaul capacity is limited to communication links between the CPU and the RAUs in cell-free networks~\cite{femenias2019cell,masoumi2019performance}.

	In this paper, we are aiming at studying the performance of NAFD cell-free mmWave massive MIMO systems, where the fronthaul link capacity is limited and hybrid analog-digital structures with low-resolution DACs are used. The main contributions of this paper are summarized as follows:
	
	
	\begin{itemize}
		\item To the best of the authors' knowledge, the problem of maximizing the weighted uplink and downlink sum rate of NAFD cell-free mmWave massive MIMO systems with fronthaul compression and DAC quantization has not been investigated. We jointly optimize the variances of the fronthaul compression noises and the power of T-RAUs and uplink users under the constraints of the bidirectional fronthaul capacity and the transmit power budgets of the T-RAUs and uplink users.
		
		\item We propose an iterative successive convex approximation based (SCA-based) algorithm to deal with the non-convex optimization problem. Specifically, the SCA method is used to convert the non-convex expressions including the optimization objective function and the uplink and downlink limited-capacity fronthaul constraints into convex forms.
		
		
	\end{itemize}
	
	The remainder of this paper is organized as follows. Section II presents the channel model, the transmission model and the problem formulation. Section III gives the proposed SCA-based algorithm to solve the optimization problem. In section IV, the simulation results are presented and analyzed. Section V concludes this paper.
	
	\textbf{Notations}: $a$, $\mathbf{a}$, $\mathbf{A}$ denotes a scalar, vector, and matrix. $\mathbb{C}^{m\times n}$ denotes the space of $m\times n$ complex matrices. $\operatorname{Tr}(\mathbf{A})$, $\operatorname{det}(\mathbf{A})$, $\mathbf{A}^{\mathrm{H}}$, $\mathbf{A}^{-1}$, respectively, denote the trace, determinant, conjugate transpose, inverse of matrix $\mathbf{A}$. $\mathbf{I}$ denotes an identity matrix whose dimensions are known from the context in a straightforward manner. The Kronecker product of $\mathbf{A}$ and $\mathbf{B}$ is $\mathbf{A}\otimes \mathbf{B}$. diag($\mathbf{A}_1$, \ldots, $\mathbf{A}_n$) denotes the block diagonal matrix which contains $\mathbf{A}_1$, \ldots, $\mathbf{A}_n$ as its block diagonal elements. vec($\mathbf{A}$) denotes the vector that stacks all the columns of matrix $\mathbf{A}$. $||\cdot||$ and $| \cdot |$ denote the Euclidean norm of a complex vector and the modulus value of a complex scalar, respectively.

	\section{System Model}
	In this section, we derive the uplink and downlink signal transmission models of NAFD cell-free mmWave massive MIMO systems by taking DAC quantization and limited-capacity fronthaul into account. We consider a scenario with $N_T$ T-RAUs, $N_R$ R-RAUs, $K$ downlink users and $J$ uplink users, where each RAU is deployed with a uniform linear antenna array (ULA) with $N_{RF}$ RF chains and $M$ antennas while each user has only single antenna. An example is given in Fig. 1.

   \begin{figure}[!h]
  \centering
  \includegraphics[width=9cm]{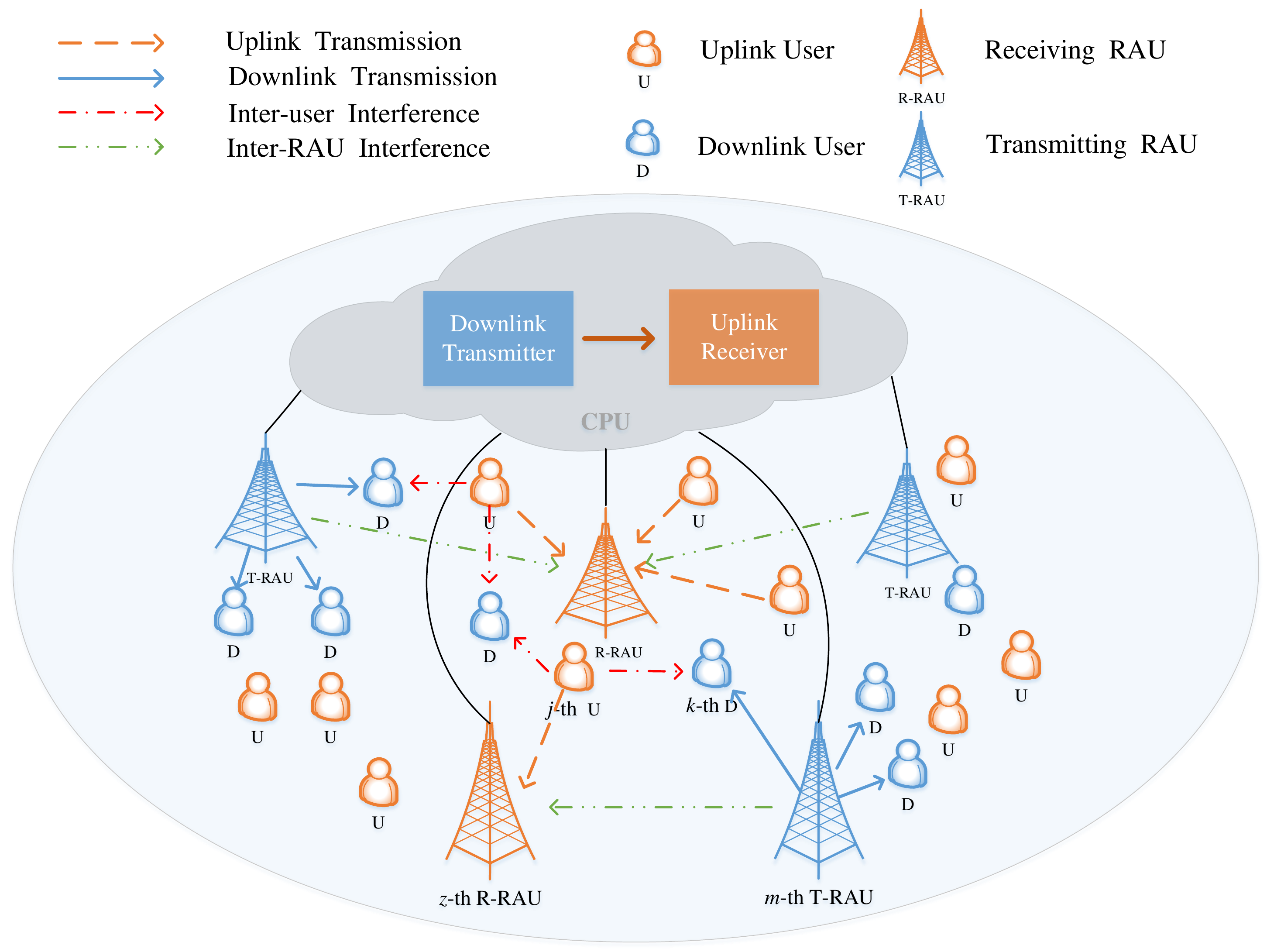}
  \caption{{System model: An NAFD cell-free mmWave massive MIMO system.}\label{system}}
\end{figure}
	
	\subsection{Channel Model}
	In this subsection, we consider an isotropic $L$-path mmWave channel model. The channel vector between the $m$-th T-RAU and the $k$-th downlink user is modeled as
	\begin{align}\label{}
		\mathbf{h}_{k, m}^\mathrm{H}=\sum_{l=1}^{L} \alpha_{k, m, l} \mathbf{v}_{rau}^\mathrm{H}\left(\theta_{k, m, l}^{\text {rau }}\right)\in\mathbb{C}^{1\times M},
	\end{align}
	where $\alpha_{k, m, l}\sim \mathcal{CN}(0,\beta_{k,m})$ and $\theta_{k, m, l}^{\text {rau }}\in [-\pi,\pi]$ are the complex gain and angle of departure (AoD) of the $l$-th path, respectively, $\beta_{k,m}$ is the large-scale fading coefficient, and the array response vector at T-RAU is defined as
	\begin{small}
			\begin{align}\label{RAU}
				\mathbf{v}_{\mathrm{rau}}\left(\theta_{k, m, l}^{\mathrm{rau}}\right) \triangleq \frac{1}{\sqrt{M}}\big[1, e^{j \kappa  d \sin (\theta_{k, m, l}^{\mathrm{rau}})}, ... , e^{j \kappa  d (M-1) \sin (\theta_{k, m, l}^{\mathrm{rau}})}\big]^{T},
			\end{align}
	\end{small}where $\kappa =\frac{2 \pi}{\lambda_{c}}$ and $d=\frac{\lambda_{c}}{2}$ is the antenna spacing. The interference channel between the $j$-th uplink user and the $k$-th downlink user is represented as
	\begin{align}\label{}
		t_{k, j} \sim \mathcal{CN}\left(0, \beta_{k, j}\right),
	\end{align}
	where $\beta_{k, j}$ is the large-scale fading coefficient. The channel vector between the $j$-th uplink user and the $z$-th R-RAU can be modeled as
	\begin{align}\label{}
		\mathbf{g}_{j, z}=\sum_{l=1}^{L} \alpha_{j, z, l} \mathbf{v}_{r a u}\left(\theta_{j, z, l}^{\text {rau }}\right)\in\mathbb{C}^{M\times 1},    	
	\end{align}
	where $\alpha_{j, z, l}\sim \mathcal{CN}(0,\beta_{j, z})$ and $\theta_{j, z, l}^{\text {rau }}\in[-\pi,\pi]$ are the complex gain and angle of arrival (AoA) of the $l$-th path, respectively, $\beta_{j, z}$ is the large-scale fading coefficient and the array response vector at R-RAU is $\mathbf{v}_{r a u}(\theta_{j, z, l}^{\text {rau }})$ with its form similar to \eqref{RAU}.
	The interference channel matrix between the $m$-th T-RAU and the $z$-th R-RAU is expressed as
	\begin{align}\label{}
		\mathbf{H}_{m, z}=\sum_{l=1}^{L} \alpha_{m, z, l} \mathbf{v}_{r a u}\left(\theta_{m, z, l}^{r}\right) &\mathbf{v}_{r a u}^{\mathrm{H}}\left(\theta_{m, z, l}^{t}\right)\in\mathbb{C}^{M\times M},
	\end{align}
	where $\alpha_{m, z, l}\sim \mathcal{CN}(0,\beta_{m, z})$ and $\beta_{m, z}$ are the complex gain of the $l$-th path and the large-scale fading coefficient, respectively,
	$\mathbf{v}_{r a u}\left(\theta_{m, z, l}^{r}\right)$ and $\mathbf{v}_{r a u}\left(\theta_{m, z, l}^{t}\right)$ are the array response vector at R-RAU and T-RAU with their forms similar to \eqref{RAU},  $\theta_{m, z, l}^{r}\in[-\pi,\pi]$ and $\theta_{m, z, l}^{t}\in[-\pi,\pi]$ are the AoA and AoD of the $l$-th path, respectively.
	
	\subsection{Transmission Model}
	As shown in Fig. \ref{system}, the $m$-th T-RAU transmits the signal to the $k$-th downlink user and the $j$-th uplink user uploads the signal to the $z$-th R-RAU simultaneously.
	
	In the downlink, the transmitted signal $\mathbf{x}_{m}\in\mathbb{C}^{M\times1}$ of the $m$-th T-RAU admits the following form
	\begin{align}\label{xm}
		\mathbf{x}_{m}=\mathbf{W}_{m} \mathrm{Q}(\sum_{{\color{blue}i=1}}^{K} \mathbf{f}_{m,{\color{blue}i}} \sqrt{\eta_{m, {\color{blue}i}}} s_{{\color{blue}i}}+\mathbf{d}_{D,m}),
	\end{align}
	where $s_{{\color{blue}i}}\sim\mathcal{CN}(0,1)$ is the data symbol transmitted to the ${\color{blue}i}$-th downlink user, $\mathbf{f}_{m, {\color{blue}i}}\in\mathbb{C}^{N_{RF}\times1}$ and $\eta_{m, {\color{blue}i}}\geq0$ are the digital precoder and power coefficient, $\mathbf{W}_{m}\in\mathbb{C}^{M\times N_{RF}}$ is the analog precoder composed of constant-modulus phase shifters at the $m$-th T-RAU, $\mathbf{d}_{D,m} \sim  \mathcal{CN}(\mathbf{0},\sigma _{D,m}^2\mathbf{I})$ is the downlink fronthaul compression noise, where $\sigma _{D,m}^2$ represents downlink fronthaul compression noise power.
	
	The quantization function $\mathrm{Q}(\text{·})$ works on the $B$-bit DACs at the RF chains of RAUs and achieves the minimum mean square error on the Gaussian signals through quantizing the real and imaginary parts with $2^B$ intervals uniformly.
	$s_{\color{blue}i}$ and $\mathbf{d}_{D,m}$ are assumed to be Gaussian distribution. The additive quantization noise model (AQNM) \cite{mezghani2012capacity} model is adopted into $\mathbf{x}_{m}$, which is equivalent to linearize the nonlinear distortion applied to Gaussian signals. Then (\ref{xm}) is rewritten as
		\begin{align}\label{xm1}
			\mathbf{x}_{m} =&\mathbf{W}_{m}\left((1-\rho)\left(\sum_{{\color{blue}i=1}}^{K} \mathbf{f}_{m, {\color{blue}i}} \sqrt{\eta_{m, {\color{blue}i}}} s_{{\color{blue}i}}+\mathbf{d}_{D,m}\right)+\overline{\mathbf{q}}_{m}\right) \nonumber\\
			=&\mathbf{W}_{m}\left(\sum_{{\color{blue}i=1}}^{K}(1-\rho) \mathbf{f}_{m, {\color{blue}i}} \sqrt{\eta_{m, {\color{blue}i}}} s_{{\color{blue}i}}+\mathbf{q}_{m}\right),
		\end{align}
	where $\overline{\mathbf{q}}_{m}\in\mathbb{C}^{N_{RF}\times 1}$ is the quantization noise, $\mathbf{q}_{m}=\overline{\mathbf{q}}_{m}+(1-\rho)\mathbf{d}_{D,m}\in\mathbb{C}^{N_{RF}\times1}$ is the total noise irrelevant to $s_{\color{blue}i}$, $\rho\textless 1$ is the quantization distortion factor related to DAC resolution. The covariance matrix of $\overline{\mathbf{q}}_{m}$ is given by \cite{mezghani2012capacity}
	\begin{align}
		\mathbf{C}_{\overline{\mathbf{q}}_{m}}=\rho(1-\rho) \operatorname{diag}\left(\sum_{{\color{blue}i=1}}^{K} \mathbf{f}_{m, {\color{blue}i}} \mathbf{f}_{m, {\color{blue}i}}^{\mathrm{H}} \eta_{m, {\color{blue}i}}+\sigma_{D,m}^{2} \mathbf{I}\right),
	\end{align}
	and the covariance matrix of $\mathbf{q}_{m}$ can be written as
	\begin{align}\label{Cq}
		\mathbf{C}_{\mathbf{q}_{m}}=\rho(1-\rho) \operatorname{diag}\left(\sum_{{\color{blue}i=1}}^{K} \mathbf{f}_{m, {\color{blue}i}} \mathbf{f}_{m, {\color{blue}i}}^{\mathrm{H}} \eta_{m, {\color{blue}i}}\right)+(1-\rho) \sigma_{D,m}^{2} \mathbf{I}.
	\end{align}
	However, the distribution of $\mathbf{q}_{m}$ is unknown  due to the non-Gaussian distribution characteristic of $\overline{\mathbf{q}}_{m}$.
	
	\setcounter{TempEqCnt}{\value{equation}}
    \setcounter{equation}{22}
	\begin{figure*}[hb]
	\hrulefill
			\begin{align}
				\left.C_{U,z}=\log _2 \Big(\operatorname{det}\big(\sum_{j=1}^{J} \mathbf{U}_{z}^\mathrm{H} \mathbf{g}_{j, z} P_{U,j} \mathbf{g}_{j, z}^\mathrm{H} \mathbf{U}_{z}+\sum_{m=1}^{N_T} \sigma_{m,z}^2 P_{D,m} \mathbf{U}_{z}^\mathrm{H} \mathbf{U}_{z}+\sigma_z^2 \mathbf{U}_{z}^\mathrm{H} \mathbf{U}_{z}+\sigma_{U,z}^{2} \mathbf{I}_{N_{RF}}\big) \Big/ \operatorname{det}\big(\sigma_{U,z}^{2} \mathbf{I}_{N_{RF}}\big)\Big) \leq C_U\right.,
			\end{align}
     \end{figure*}
     \setcounter{equation}{\value{TempEqCnt}}
	The received signal of the $k$-th downlink user is
		\begin{align}\label{yd,k}
			y_{D,k}=\sum_{m=1}^{N_T} \mathbf{h}_{k, m}^\mathrm{H} \mathbf{x}_{m}+\sum_{j=1}^{J} t_{k, j} \sqrt{P_{U, j}} s_{U, j}+n_{D, k},
		\end{align}
	where {\color{red}the subscript `$D$' and `$U$' mean downlink and uplink respectively, the subscript `$k$' and `$j$' correspond to the $k$-th downlink user and the $j$-th uplink user rsespectively}, $s_{U,j}$ is the data symbol transmitted by the $j$-th uplink user, $P_{U,j}$ is the transmit power of the $j$-th uplink user, $n_{D,k}\sim\mathcal{CN}(0,\sigma _{k}^2)$ is the additive white Gaussian noise with its variance $\sigma _{k}^2$. Substituting (\ref{xm1}) into (\ref{yd,k}), (\ref{yd,k}) can be rewritten as
		\begin{align}
			y_{D,k}=&\sum_{m=1}^{N_T} \mathbf{h}_{k, m}^\mathrm{H}\mathbf{W}_{m}\left(\sum_{{\color{blue}i=1}}^{K}(1-\rho) \mathbf{f}_{m, {\color{blue}i}} \sqrt{\eta_{m, {\color{blue}i}}} s_{{\color{blue}i}}+\mathbf{q}_{m}\right)\nonumber\\
			&{+\sum_{j=1}^{J} t_{k, j} \sqrt{P_{U, j}} s_{U, j}+{n_{D, k}}.}
		\end{align}
	To simplify the above expression, let $\overline{\mathbf{h}}_{k, m}=\mathbf{W}_{m}^{\mathrm{H}} \mathbf{h}_{k, m}$, $\mathbf{q}^{\mathrm{H}}=\left[\begin{array}{lll}\mathbf{q}_{1}^{\mathrm{H}}  \cdots  \mathbf{q}_{N_T}^{\mathrm{H}}\end{array}\right]$, $\mathbf{h}_{k}^{\mathrm{H}}=\left[\begin{array}{lll}\overline{\mathbf{h}}_{k, 1}^{\mathrm{H}}  \cdots  \overline{\mathbf{h}}_{k, N_T}^{\mathrm{H}}
	\end{array}\right]$. Thus, $y_{D,k}$ can be transformed into the following form
		\begin{align}\label{yd,k1}
			y_{D,k} =&\sum_{m=1}^{N_T} \sum_{{\color{blue}i=1}}^{K}(1-\rho) \overline{\mathbf{h}}_{k, m}^{\mathrm{H}} \mathbf{f}_{m, {\color{blue}i}} \sqrt{\eta_{m, {\color{blue}i}}} s_{{\color{blue}i}}+\sum_{m=1}^{N_T} \overline{\mathbf{h}}_{k, m}^{\mathrm{H}} \mathbf{q}_{m}\nonumber \\
			&+\sum_{j=1}^{J} t_{k, j} \sqrt{P_{U, j}} s_{U, j}+ n_{D,k}\nonumber\\
			=&\sum_{m=1}^{N_T}(1-\rho) \overline{\mathbf{h}}_{k, m}^{\mathrm{H}}\mathbf{f}_{m, k} \sqrt{\eta_{m, k}} s_{k}+\sum_{i \neq k} \sum_{m=1}^{N_T}(1-\rho) \overline{\mathbf{h}}_{k, m}^{\mathrm{H}}\nonumber\\
			&\times \mathbf{f}_{m, i}\sqrt{\eta_{m, i}} s_{i}+\mathbf{h}_{k}^{\mathrm{H}} \mathbf{q}+\sum_{j=1}^{J} t_{k, j} \sqrt{P_{U, j}} s_{U,j}+ n_{D,k}.
		\end{align}
	
	To eliminate the IUI in the second term in (\ref{yd,k1}), the ZF precoder is used at the basedband of T-RAUs. {\color{blue}The power coefficient $\eta_{m, k}$ is forced to only depend on $k$, so the subscript $m$ is removed from $\eta_{m,k}$, which results in $\bm{\eta}=\operatorname{diag}\left(\eta_{1}, \ldots, \eta_{K}\right)$} {\color{red}\cite{kim2021performance}}.
	Let $\mathbf{F}_{m}=\left[\begin{array}{lll}\mathbf{f}_{m, 1} & \cdots & \mathbf{f}_{m, K}\end{array}\right], \mathbf{F}^{\mathrm{H}}=\left[\begin{array}{lll}\mathbf{F}_{1}^{\mathrm{H}} & \cdots & \mathbf{F}_{N_T}^{\mathrm{H}}\end{array}\right]$, $\mathbf{H}^{\mathrm{H}}=\left[\begin{array}{lll}
		\mathbf{h}_{1} & \cdots & \mathbf{h}_{K}
	\end{array}\right]$.
	The ZF precoder should satisfy the following condition
	\begin{equation}\label{}
		\sum_{m=1}^{N_T} \overline{\mathbf{h}}_{k, m}^{\mathrm{H}} \mathbf{f}_{m, i}=\left\{\begin{array}{l}
			1 \text { if } i=k \\
			0 \text { if } i \neq k
		\end{array}\right.,
	\end{equation}
	which can be realized by setting $\mathbf{F}$ as the pseudo-inverse of $\mathbf{H}$. Through interference cancellation, the received signal of the $k$-th downlink user becomes
	\begin{align}\label{}
		\overline{y}_{D,k}=&(1-\rho) \sqrt{\eta_{k}} s_{k}+\mathbf{h}_{k}^\mathrm{H} \mathbf{q}+\sum_{j=1}^{J}  t_{k, j} \sqrt{P_{U, j}} s_{U, j}\nonumber\\
		&+n_{D,k}.
	\end{align}
	
	Since the distribution of $\mathbf{q}$ is unknown, the closed-form rate expression of the $k$-th downlink user is difficult to obtain. Therefore, with the worst noise situation considered, the lower bound of the achievable rate of downlink users can be obtained \cite{mezghani2012capacity}. Since the worst-case noise in the distribution with the same covariance matrix is Gaussian distribution and the covariance matrix of $\mathbf{q}$ is represented as
		\begin{align}\label{CQ}
			\mathbf{C}_{\mathbf{q}} &=\operatorname{diag}\left(\mathbf{C}_{\mathbf{q}_{1}}, \ldots, \mathbf{C}_{\mathbf{q}_{N_T}}\right) \nonumber \\
			&=\rho(1-\rho) \operatorname{diag}\left(\mathbf{F} \bm{\eta} \mathbf{F}^{\mathrm{H}}\right)+(1-\rho) \bm{\sigma}_D^{2} \otimes \mathbf{I},
		\end{align}	
	where $\bm{\eta}=\operatorname{diag}\left(\eta_{1}, \ldots, \eta_{K}\right)$, $\bm{\sigma}_D=\operatorname{diag}\left(\sigma_{D,1}, \ldots, \sigma_{D,N_T}\right)$.
	{\color{blue}After that, the lower bound for the achievable rate of the $k$-th downlink user is derived as} {\color{red}\cite{kim2021performance}}
	\begin{small}
	\begin{align}\label{}
		R_{D,k} = \log_2(1+\underbrace{\frac{(1-\rho)^2\eta_{k}}{\mathbf{h}_k^\mathrm{H}\mathbf{C}_{\mathbf{q}}\mathbf{h}_k+\sum_{j=1}^{J}\left|t_{k, j} \sqrt{P_{U, j}}\right|^{2}+\sigma_{k}^{2}}}_{=\tilde{\gamma}_k}).
	\end{align}
	\end{small}
	
	According to the rate distortion theory, when conveying the compressed transmit signal from CPU to T-RAU, the fronthaul capacity requirement must be satisfied, which is expressed as \cite{yang2021joint}
	
	\begin{align}
		C_{D,m}=&\log _{2}\left(\frac{\operatorname{det}\big(\sum_{k=1}^{K} \mathbf{f}_{m, k} \mathbf{f}_{m, k}^\mathrm{H} \eta_{k}+\sigma_{D, m}^{2} \mathbf{I}_{N_{R F}}\big)}{\operatorname{det}\big(\sigma_{D, m}^{2} \mathbf{I}_{N_{R F}}\big)}\right)\nonumber \\
		\leq& C_D,
		\end{align}
	where $C_D$ represents the fronthaul compression maximum capacity.
	
	In the uplink, the received signal of the $z$-th R-RAU is
	\begin{align}\label{}
		y_{U,z}=\mathbf{U}_{z}^\mathrm{H}\left(\sum_{j=1}^{J} \mathbf{g}_{j, z} \sqrt{P_{U, j}} s_{U, j}+\sum_{m=1}^{N_T} \mathbf{H}_{m, z} \mathbf{x}_{m}+\mathbf{n}_{U,z}\right),
	\end{align}
	where $\mathbf{U}_{z}\in\mathbb{C}^{M\times N_{RF}}$ is the analog receiver of the $z$-th R-RAU which also has constant modulus constraints, $\mathbf{n}_{U,z}\sim \mathcal{CN}(\bm{0},\sigma_{z}^2\mathbf{I})\in \mathbb{C}^{M\times 1}$ is the additive white Gaussian noise with its {\color{red}elements} variance {\color{blue}$\sigma_{z}^2$}. After the IRI cancellation performed by the CPU in the digital domain~\cite{wang2019performance}, the received signal is expressed as
		\begin{align}
			\tilde{y}_{U,z}=&\mathbf{U}_{z}^\mathrm{H}\left(\sum_{j=1}^{J} \mathbf{g}_{j, z} \sqrt{P_{U, j}} s_{U, j}+\sum_{m=1}^{N_T} \mathbf{\widetilde{H}}_{m, z} \mathbf{x}_{m}+\mathbf{n}_{U,z}\right) \nonumber\\
			&+\mathbf{d}_{U,z},
		\end{align}
	where $\mathbf{d}_{U,z}\sim\mathcal{CN}(\mathbf{0},\sigma_{U,z}^2\mathbf{I})\in\mathbb{C}^{N_{RF}\times 1}$ is the fronthaul compression noise, $\mathbf{\widetilde{H}}_{m, z}$ represents the channel estimation error. It is assumed that the elements of $\mathbf{\widetilde{H}}_{m, z}$ follow the independent identically distributed Gaussian distribution, i.e.,$\mathrm{vec}(\mathbf{\widetilde{H}}_{m, z}) \sim \mathcal{C} \mathcal{N}\left(\mathbf{0}, \sigma_{m,z}^{2} \mathbf{I}_{M\times M}\right)$, where $\sigma_{m,z}^{2}$ represents the residual interference power~\cite{yang2021joint}.
    \setcounter{TempEqCnt}{\value{equation}}
    \setcounter{equation}{28}
    \begin{figure*}[hb]
	\hrulefill
	\begin{align}\label{yueshu44}
		&\log \Big( \operatorname{det}\big(\mathbf{B}_{t}\big)\Big)+\sum_{j=1}^{J} \mathbf{g}_{j, z}^{\mathrm{H}} \mathbf{U}_{z} \mathbf{B}_{t}^{-1} \mathbf{U}_{z}^{\mathrm{H}} \mathbf{g}_{j, z} P_{U, j}+\sum_{m=1}^{N_T} \sigma_{m,z}^2 P_{D,m} \operatorname{Tr}(\mathbf{U}_{z}^{\mathrm{H}} \mathbf{U}_{z} \mathbf{B}_{t}^{-1})\nonumber\\
		&+\sigma_z^2\operatorname{Tr}(\mathbf{U}_{z}\mathbf{B}_{t}^{-1}\mathbf{U}_{z}^{\mathrm{H}})+\sigma_{U,z}^{2}\operatorname{Tr}(\mathbf{B}_{t}^{-1})-N_{R F}-2N_{RF}\log (\sigma_{U, z}) \leq C_{U} \log (2),
	\end{align}
    \end{figure*}
    \setcounter{equation}{\value{TempEqCnt}}

	To eliminate the interference from the other uplink users, a ZF receiver is used at the R-RAU. Denoting the digital receiving matrix of the $z$-th R-RAU at the baseband by $\mathbf{v}_z\in\mathbb{C}^{N_{RF}\times 1}$. $\mathbf{v}_z$ is designed as $\mathbf{v}_z=\overline{\mathbf{g}}_{j,z}(\overline{\mathbf{g}}_{j,z}^\mathrm{H}\overline{\mathbf{g}}_{j,z})^{-1}$, where $\overline{\mathbf{g}}_{j,z}=\mathbf{U}_z^\mathrm{H}\mathbf{g}_{j, z}\in\mathbb{C}^{N_{RF}\times 1}$.
	Let $\mathbf{G}_{m, z}=\mathbf{W}_{m}^\mathrm{H} \mathbf{\widetilde{H}}_{m, z}^\mathrm{H} \mathbf{U}_{z}, \mathbf{G}_{z}^\mathrm{H}=\left[\mathbf{G}_{1, z}^\mathrm{H} \ldots \mathbf{G}_{N_T, z}^\mathrm{H}\right]$. After the IUI cancellation, the received signal becomes
		\begin{align}
			\overline{y}_{U,z}=&\mathbf{v}_{z}^\mathrm{H}(\overline{\mathbf{g}}_{j,z} \sqrt{P_{U, j}} s_{U, j}+\sum_{m=1}^{N_T} \sum_{k=1}^{K} (1-\rho)\mathbf{G}_{m, z}^{\mathrm{H}} \mathbf{f}_{m, k} \sqrt{\eta_{k}} s_{k} \nonumber\\
			&+\mathbf{G}_{z}^{\mathrm{H}} \mathbf{q}+\mathbf{U}_{z}^{\mathrm{H}} \mathbf{n}_{U,z}+\mathbf{d}_{U,z}).
		\end{align}
	
	Therefore, {\color{blue}the lower bound for the achievable rate} of the $j$-th uplink user can be written as
	\begin{equation}\label{}
		\begin{aligned}
			R_{U,j}=\log_{2}(1+\gamma_j),
		\end{aligned}
	\end{equation}
	where
	\begin{equation}\label{}
		\gamma_j=\underbrace{\frac{P_{U, j}|\mathbf{v}_{z}^\mathrm{H} \overline{\mathbf{g}}_{j,z}|^{2}}
			{A+B+C+D}}_{=\tilde{\gamma}_j},
	\end{equation}
	$A=(1-\rho)^{2} \sum^{N_T}_{m=1} \sum_{k=1}^{K} \eta_{k}|\mathbf{v}_{z}^\mathrm{H} \mathbf{G}_{m, z}^\mathrm{H} \mathbf{f}_{m, k}|^{2}, B=\mathbf{v}_{z}^\mathrm{H}\mathbf{G}_{z}^\mathrm{H} \mathbf{C}_{q} \mathbf{G}_{z}\mathbf{v}_{z}, C=\sigma_z^2||\mathbf{v}_{z}^\mathrm{H} \mathbf{U}_{z}^\mathrm{H} ||^{2}, D=\sigma_{U, z}^{2}||\mathbf{v}_{z}||^{2}$.
	
	Similar to the downlink transmission, when conveying the compressed received signal $y_{U,z}$ from R-RAU to the CPU, the fronthaul capacity constraint is shown at the bottom of this page, where $C_U$ represents the fronthaul compression maximum capacity.
	
	\subsection{Problem Formulation}
	In this subsection, we propose to maximize the weighted uplink and downlink sum rate under the constraints of fronthaul capacity, transmit power of RAUs and uplink users. The set of variables to be optimized is denoted as $\mathcal{A}_0\triangleq \left\{\sigma_{D, m}^2,\sigma_{U, z}^2,\eta_{k},P_{U,j}\right\}$. Then, the mathematical expression of the optimization problem can be written as
    \setcounter{equation}{23}
    \begin{subequations}
			\begin{alignat}{2}
			&\max _{\mathcal{A}_0} \quad& &\omega _D \sum_{k=1}^{K} R_{D,k}+ \omega_U \sum_{j=1}^{J} R_{U,j}\label{mubiao} \\
			& \;\; \text { s.t. }
			& \quad & P_{D,m}\leq P_D ,  \quad \forall m, \label{yueshu1} \\
			&&& C_{D,m}\leq C_D,  \quad \forall m, \label{yueshu2}\\
			&&& P_{U,j}\leq P_U, \quad  \forall j, \label{yueshu3}\\
			&&&  C_{U,z}\leq C_U, \quad \forall z, \label{yueshu4}
		\end{alignat}
	\end{subequations}
	where $P_{D,m}$ {\color{blue}is the power of the precoded and quantized signal $\mathbf{x}_m$ transmitted by the $m$-th T-RAU, $P_{U,j}$ is the power of the $j$-th uplink user}. $P_D$ and $P_U$ are the maximum transmit power of T-RAUs and uplink users, respectively. $\omega_D$ and $\omega _U$ are the weighted coefficients of downlink and uplink sum rate, respectively, which satisfy $\omega_D+\omega_U=1$ and ensure the fairness of communication. (\ref{yueshu1}) represents the power constraint of T-RAUs, (\ref{yueshu3}) constrains the transmit power of uplink users, (\ref{yueshu2}) and (\ref{yueshu4}) restrict the downlink and uplink fronthaul compression capacity.
	To satisfy the constant modulus constraints of $\mathbf{W}_m$ and $\mathbf{U}_z$, the alternate projection method~\cite{tropp2005designing,ribeiro2018energy} is adopted.
    Since (\ref{yueshu2}) and (\ref{yueshu4}) are non-convex, we turn them into the convex ones by using the SCA method in the next section.
	
	\section{Proposed Iterative SCA-based Algorithm}
	Considering that $\mathbf{q}_m$ is irrelevant to $s_k$, the transmit power of the $m$-th T-RAU can be expressed in a matrix form {\color{red}by its definition $P_{D,m} = \mathbb{E}[\Vert \mathbf{x}_m \Vert^2]=\mathbb{E}[\operatorname{Tr}(\mathbf{x}_m\mathbf{x}_m^{\mathrm{H}})]$} as
		\begin{align}
			P_{D,m}\left(\boldsymbol{\eta}, \sigma_{D,m}\right)=&(1-\rho)^{2} \operatorname{Tr}\left(\mathbf{W}_{m} \mathbf{F}_{m} \boldsymbol{\eta} \mathbf{F}_{m}^{\mathrm{H}} \mathbf{W}_{m}^{\mathrm{H}}\right) \nonumber \\
			&+\rho(1-\rho) \operatorname{Tr}\left(\mathbf{W}_{m} \operatorname{diag}\left(\mathbf{F}_{m} \boldsymbol{\eta} \mathbf{F}_{m}^{\mathrm{H}}\right) \mathbf{W}_{m}^{\mathrm{H}}\right) \nonumber\\
			&+(1-\rho) \sigma_{D,m}^{2} \operatorname{Tr}\left(\mathbf{W}_{m} \mathbf{W}_{m}^{\mathrm{H}}\right).
		\end{align}
	
	The downlink fronthaul capacity constraint $\left\{C_{D,m}\leq C_D,  \forall m\right\}$ can be rewritten as
		\begin{align}\label{yueshu2_}
\log\Big( \operatorname{det}\big(&\sum\nolimits_{k=1}^{K} \mathbf{f}_{m, k} \mathbf{f}_{m, k}^\mathrm{H} \eta_{k}+\sigma_{D, m}^{2} \mathbf{I}_{N_{R F}}\big)\Big) \nonumber\\
& -\log\Big( \operatorname{det}\big(\sigma_{D, m}^{2} \mathbf{I}_{N_{R F}}\big)\Big)\leq C_D \log(2).
		\end{align}
    Using the first-order Taylor expansion for $\log(\operatorname{det}(\boldsymbol{X}))$ \cite{boyd2004convex}
    \begin{align}
        \log (\operatorname{det}(\boldsymbol{X})) \leq \log \left(\operatorname{det}\left(\boldsymbol{X}_{0}\right)\right)+\operatorname{Tr}\left(\boldsymbol{X}_{0}^{-1}\left(\boldsymbol{X}-\boldsymbol{X}_{0}\right)\right),\nonumber
    \end{align}
    \eqref{yueshu2_} can be approximated as
		\begin{align}\label{yueshu22}
			&\sum_{k=1}^{K} \mathbf{f}_{m, k}^\mathrm{H} \mathbf{A}_{t}^{-1} \mathbf{f}_{m, k} \eta_{k}+\sigma_{D, m}^{2} \operatorname{Tr}\left(\mathbf{A}_{t}^{-1}\right) \leq N_{R F}+2N_{R F}\nonumber\\
			&\times \log \left(\sigma_{D, m}\right)+C_{D} \log (2)-\log\big( \operatorname{det} \big(\mathbf{A}_{t}\big)\big),
		\end{align}
	where
	\begin{align}\label{At}
		\mathbf{A}_{t}=\sum_{k=1}^{K} \mathbf{f}_{m, k} \mathbf{f}_{m, k}^\mathrm{H} \eta_{k}^{(t)}+\left(\sigma_{D, m}^{(t)}\right)^{2} \mathbf{I}_{N_{R F}}.
	\end{align}
	
	Similar to the downlink scenario, the uplink fronthaul capacity constraint $\left\{C_{U,z}\leq C_U,  \forall z\right\}$ can be approximated as (\ref{yueshu44}) shown at the bottom of this page, where
	\setcounter{equation}{29}
		\begin{align}\label{Bt}
			\mathbf{B}_{t}=&\sum_{j=1}^{J} \mathbf{U}_{z}^{\mathrm{H}} \mathbf{g}_{j,z} P_{U,j}^{(t)} \mathbf{g}_{j,z}^{\mathrm{H}} \mathbf{U}_z+\sum_{m=1}^{N_T} \sigma_{m,z}^2 P_{D,m}^{(t)} \mathbf{U}_z^{\mathrm{H}} \mathbf{U}_z \nonumber\\
			&+\sigma_z^2 \mathbf{U}_z^{\mathrm{H}} \mathbf{U}_z+(\sigma_{U, z}^{(t)})^{2} \mathbf{I}_{N_{RF}}.
		\end{align}
	
	Therefore, the original problem can be rewritten as
	\begin{subequations}
		\begin{alignat}{2}
			&\max _{\mathcal{A}_0} \quad& &\omega _D \sum_{k=1}^{K} R_{D,k}+ \omega_U \sum_{j=1}^{J} R_{U,j}\\
			& \;\; \text { s.t. }
			& \quad & (\ref{yueshu1}), (\ref{yueshu22}), (\ref{yueshu3}), (\ref{yueshu44}).
        \end{alignat}
	\end{subequations}
	
	In the following, we convert the objective function to a convex form. Asumming that $\omega_D=\omega_U=1/2$, the objective function can be expressed as the difference between two concave functions, i.e.,
	\begin{equation}\label{}
		\sum_{k=1}^{K} R_{D,k}+ \sum_{j=1}^{J} R_{U,j}=f(\mathcal{A}_0)-h(\mathcal{A}_0),
	\end{equation}
	where
			$f(\mathcal{A}_0)=\sum_{k=1}^{K}\log_{2}((1-\rho)^2\eta_{k}+\tilde{\gamma}_k)
			+\sum_{j=1}^J\log_{2}(P_{U,j}|\mathbf{v}_{z}^\mathrm{H} \overline{\mathbf{g}}_{j,z}|^{2}+\tilde{\gamma}_j)$,
		$h(\mathcal{A}_0)=\sum_{k=1}^K\log_{2}(\tilde{\gamma}_k)+\sum_{j=1}^J\log_{2}(\tilde{\gamma}_j)$.
	
	Inspired by \cite{kha2011fast}, a first-order approximation of concave function $h(\mathcal{A}_0)$ can be expressed as
		\begin{align}
			h^{(n)}(\mathcal{A})=&h\left(\mathcal{A}^{(n)}\right)+\frac{\phi_{D}^{(n)}}{\ln2} \sum_{k=1}^{K} \mathbf{h}_{\mathrm{k}}^{\mathrm{H}}\left(\mathbf{C}_{\mathbf{q}}-\mathbf{C}_{\mathbf{q}}^{(\mathrm{n})}\right) \mathbf{h}_{\mathrm{k}}\nonumber \\
			&+\frac{\phi_{D}^{(n)}}{\ln 2}  \sum_{k=1}^{K} \sum_{\mathrm{j}=1}^{\mathrm{J}}\left| \mathrm{t}_{\mathrm{k}, \mathrm{j}}\right|^{2}\left(\mathrm{P}_{\mathrm{U}, \mathrm{j}}-\mathrm{P}_{\mathrm{U}, \mathrm{j}}^{(\mathrm{n})}\right)\nonumber\\
			&+\frac{\phi_{U}^{(n)}}{\ln 2} \sum_{j=1}^{J} \mathbf{v}_z^\mathrm{H} \mathbf{G}_{z}^{\mathrm{H}}\left(\mathbf{C}_{\mathbf{q}}-\mathbf{C}_{\mathbf{q}}^{(\mathrm{n})}\right) \mathbf{G}_{z}\mathbf{v}_z \nonumber\\
			&+\frac{\phi_{U}^{(n)}}{\ln 2}  \sum_{j=1}^{J}\left(\sigma_{U, z}^{2}-\sigma_{U, z}^{{(n)}^2}\right)\left\|\mathbf{v}_{z}\right\|^{2}\nonumber\\
&+\frac{\phi_{U}^{(n)}}{\ln 2}  \sum_{j=1}^{J}(1-\rho)^{2} \sum_{m=1}^{N_T} \sum_{k=1}^{K}\left(\eta_{k}-\eta_{k}^{(n)}\right)\nonumber\\
&\times\left|\mathbf{v}_{z}^{\mathrm{H}} \mathbf{G}_{m, z}^{\mathrm{H}} \mathbf{f}_{m, k}\right|^{2},
		\end{align}
	where
	\begin{align}\label{fid}
		\phi_{D}^{(n)}=\Bigg[\mathbf{h}_{\mathrm{k}}^{\mathrm{H}} \mathbf{C}_{\mathbf{q}}^{(\mathrm{n})} \mathbf{h}_{\mathrm{k}}+\sigma_{ k}^{2}+\sum_{\mathrm{j}=1}^{J}\left| \mathrm{t}_{\mathrm{k}, \mathrm{j}} \sqrt{\mathrm{P}_{U, j}^{(\mathrm{n})}}\right|^{2}\Bigg]^{-1},
	\end{align}
		\begin{align}\label{fiu}
			&\phi_{U}^{(n)}=\Bigg[(1-\rho)^{2} \sum_{m=1}^{N_T} \sum_{k=1}^{K} \eta_{k}^{(n)}|\mathbf{v}_{z}^\mathrm{H} \mathbf{G}_{m, z}^{\mathrm{H}} \mathbf{f}_{m, k}|^{2}+ \nonumber\\
			&\mathbf{v}_z^\mathrm{H} \mathbf{G}_{z}^{\mathrm{H}} \mathbf{C}_{\mathbf{q}}^{(n)} \mathbf{G}_{z} \mathbf{v}_z+\sigma_z^2||\mathbf{v}_{z}^\mathrm{H} \mathbf{u}_{z}^{\mathrm{H}} ||^{2} +\sigma_{U, z}^{(n) 2}\left\|\mathbf{v}_{z}\right\|^{2}\Bigg]^{-1}.
		\end{align}
	
	Finally, we can rewrite the optimization problem at the $n+1$ iteration as
	\begin{subequations}\label{youhua}
		\begin{alignat}{2}
			&\max _{\mathcal{A}_0} \quad& &f(\mathcal{A})-h^{(n)}(\mathcal{A}) \\
			& \;\; \text { s.t. }
			& \quad & (\ref{yueshu1}), (\ref{yueshu22}), (\ref{yueshu3}), (\ref{yueshu44}).
		\end{alignat}
	\end{subequations}
	
	The solution to problem (\ref{youhua}) as the so-called iterative SCA-based algorithm is summarized in Algorithm 1. Since problem \eqref{youhua} is a linear program (LP), according to \cite{wang2014outage}, the complexity of solving this problem is $\mathcal{O}(\sqrt{2N_t+J+N_R}\cdot n\cdot[(n+1)(2N_T+J+N_R)+n^2]$, where $n=\mathcal{O}(max(N_T, N_R, K, J))$.

	\begin{algorithm}[t]
		\renewcommand{\algorithmicrequire}{\textbf{Input:}}
		\renewcommand{\algorithmicensure}{\textbf{Output:}}
		\caption{The proposed iterative SCA-based algorithm}
		\label{alg1}
		\begin{algorithmic}[1]
			\STATE Initialization: $\left\{\sigma_{D, m}^2,\sigma_{U, z}^2,\eta_{k},P_{U,j}\right\}^{(0)}$;
			\STATE set $n\rightarrow 0$;
			\STATE update$\left\{\mathbf{C}_\mathbf{q}, \mathbf{A}_t, \mathbf{B}_t, \phi_{D},\phi_{U}\right\}^{(n)}$ according to (\ref{CQ}), (\ref{At}), (\ref{Bt}), (\ref{fid}), (\ref{fiu});
			\REPEAT
			\STATE solve (\ref{youhua}) through cvx in Matlab and get values of variables at the $n+1$ iteration $\mathcal{A}_0^{(n+1)}$;
			\STATE update $\left\{\mathbf{C}_\mathbf{q}, \mathbf{A}_t, \mathbf{B}_t, \phi_{D},\phi_{U}\right\}^{(n+1)}$;
			\STATE set $n \rightarrow n+1$;
			\UNTIL $\bigg|\big[f(\mathcal{A})-h^{(n)}(\mathcal{A})\big]^{(n+1)}-\big[f(\mathcal{A})-h^{(n)}(\mathcal{A})\big]^{(n)}\bigg|\leq 10^{-3}$ is met.
		\end{algorithmic}
	\end{algorithm}

	\section{Simulation Results}
\begin{figure}
	\centering
	\includegraphics[width=3.3in]{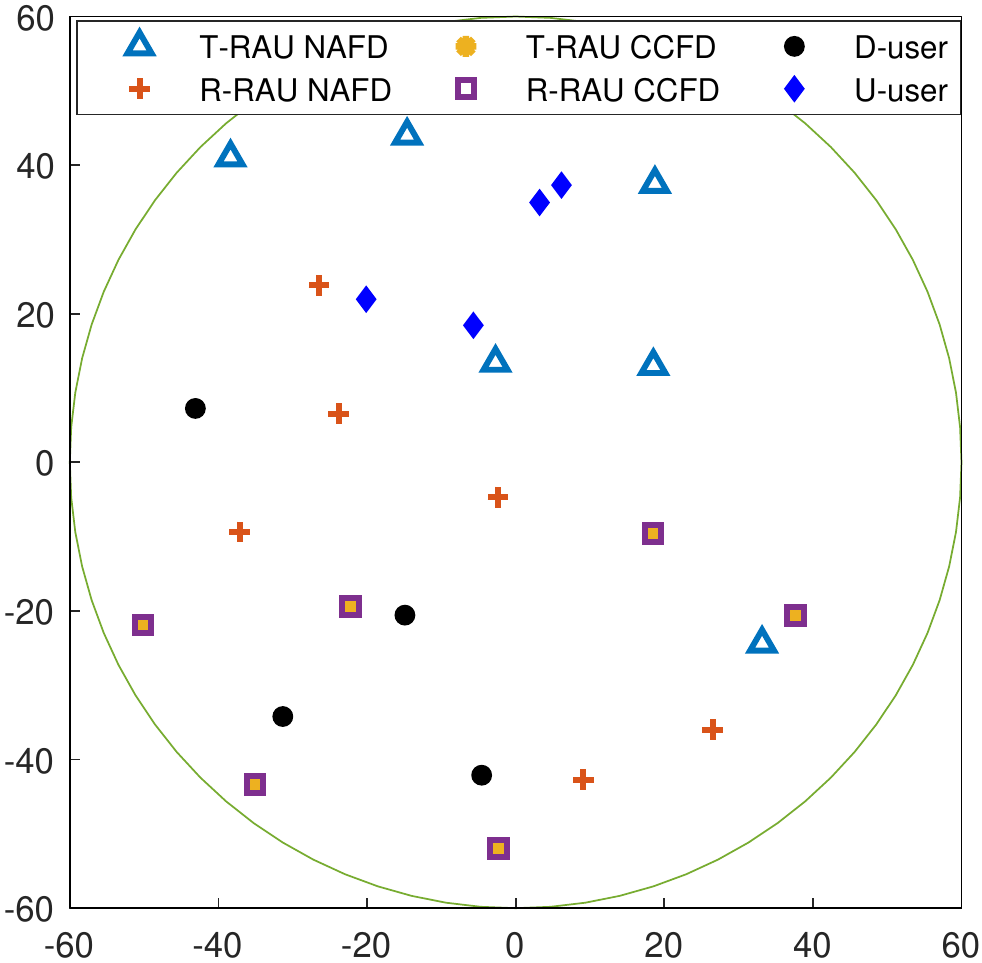}
	\caption{RAUs and users layout.}
	\label{RAUs and users layout}
	\end{figure}

	In this section, the performance of achievable sum rate in NAFD cell-free mmWave massive MIMO system with limited-capacity fronthaul links and low-resolution DACs is measured. For simplicity, both $\omega_D$ and $\omega_U$ are set to be 1/2. {\color{blue}We consider a circular area with radius $R=60$ m and all the RAUs and users are randomly distributed in this area. Specifically, all the T-RAUs and R-RAUs in NAFD system are geographically separated, while one T-RAU and one R-RAU in CCFD C-RAN system are paired and distributed in the same position. \cite{wang2019performance, li2020network}. A specific case of the above distribution is shown as Fig. \ref{RAUs and users layout}.} The protection distance between the user and the RAU is $r_0$. The {\color{red}additional} detailed simulation parameters are listed in Table \ref{table_simulation}. According to the simulation parameters, the noise power can be calculated by $\sigma_z^2=\sigma_{k}^2=-174$ dBm/Hz$+10\log_{10}(B)+$NF, $\forall k,z$.
 The large-scale fading coefficient on distance $d$ (m) is modeled as
	\begin{align}
		\mathrm{PL}(d)[\mathrm{dB}]=\mathrm{PL}(d_0)+10\xi \log _{10}(\frac{d}{d_0})+X_{\zeta}.
	\end{align}
	\begin{table}
		\caption{Simulation Parameters}
		\label{table_simulation}
		\setlength{\tabcolsep}{0.9mm}
		\centering
		\begin{tabular}{l l}
			\toprule[1.5pt]
			\textbf{Parameter} & \textbf{Value} \\
			\midrule
			{\color{blue}RAU and User distribution radius} $R$ & {\color{blue}60} m\\
			Protection distance $r_0$   &5m\\
			Carrier frequency $f_0$ & 28 GHz \\
			Bandwidth $B$ & 100 MHz \\
			Num. of T-RAUs $N_T$/R-RAUs $N_R$ & 6/6\\
			Num. of uplink users $J$/downlink users $K$  & 4/4 \\
			Num. of RAU antennas $M$/RF chains $N_{RF}$ & 6/3 \\
			Num. of channel paths $L$ &6\\
			Noise figure NF & 9 dB\\
			Noise power ${\color{blue}\sigma_{k}^{2}}$(${\color{blue}\sigma_z^2}$) & -85 dBm\\
			Residual interference power $\sigma_{m,z}^2$, $\forall m,z$ & -105 dBm\\
			T-RAUs/users transmit power $P_D$/$P_U$ & 30/27 dBm \\
			Path angle distribution & $\mathcal{U} \left( -\pi, \pi \right)$ \\
			Free space path loss PL($d_0$)[dB] & 20$\log_{10}(\frac{4\pi d_0}{\lambda_{c}})$ \\
			Path loss index $\xi$ & 2.92\\
			Logarithm shadow fading distribution $X_{\zeta}$ & $\mathcal{N}(0,8.7^2)$\\
			Reference distance $d_0$ & 1 m\\
			\bottomrule[1.5pt]
		\end{tabular}
	\end{table}
	
\begin{figure}
		\centering
		\includegraphics[width=3.3in]{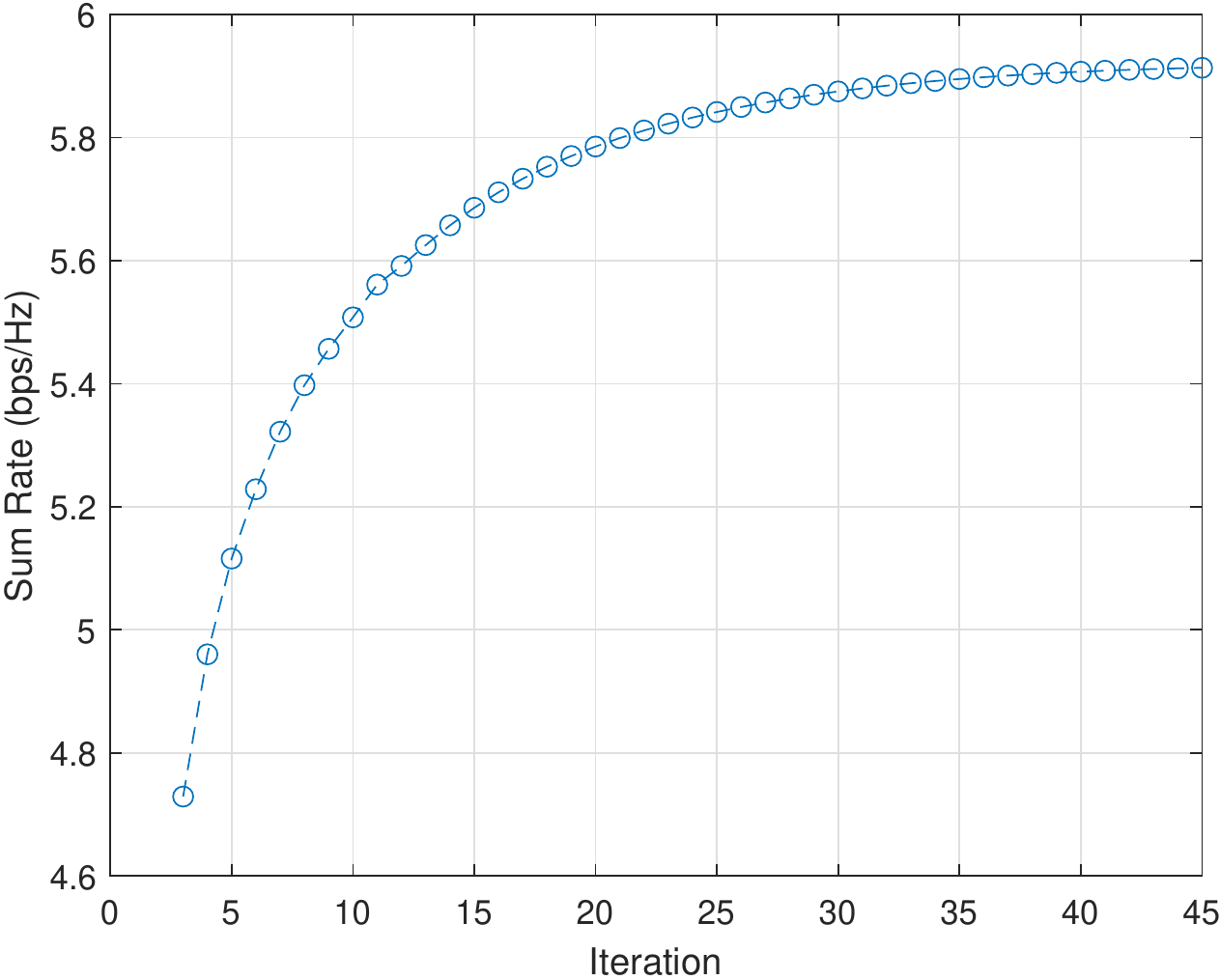}
		\caption{Convergence performance of the proposed algorithm with $B=1$ bit, $C_D=C_U=26$ bps/Hz.}
		\label{shoulian}
	    \end{figure}

	First, as shown in Fig. \ref{shoulian}, we verify the convergence of the proposed algorithm with the fronthaul compression capacity $C_D=C_U=26$ bps/Hz and $B=1$ bit. As the iteration reaches $43$, the lower bound of the achievable sum rate monotonously converges to 5.91 bps/Hz.
	
	\begin{figure}
		\centering
		\includegraphics[width=3.3in]{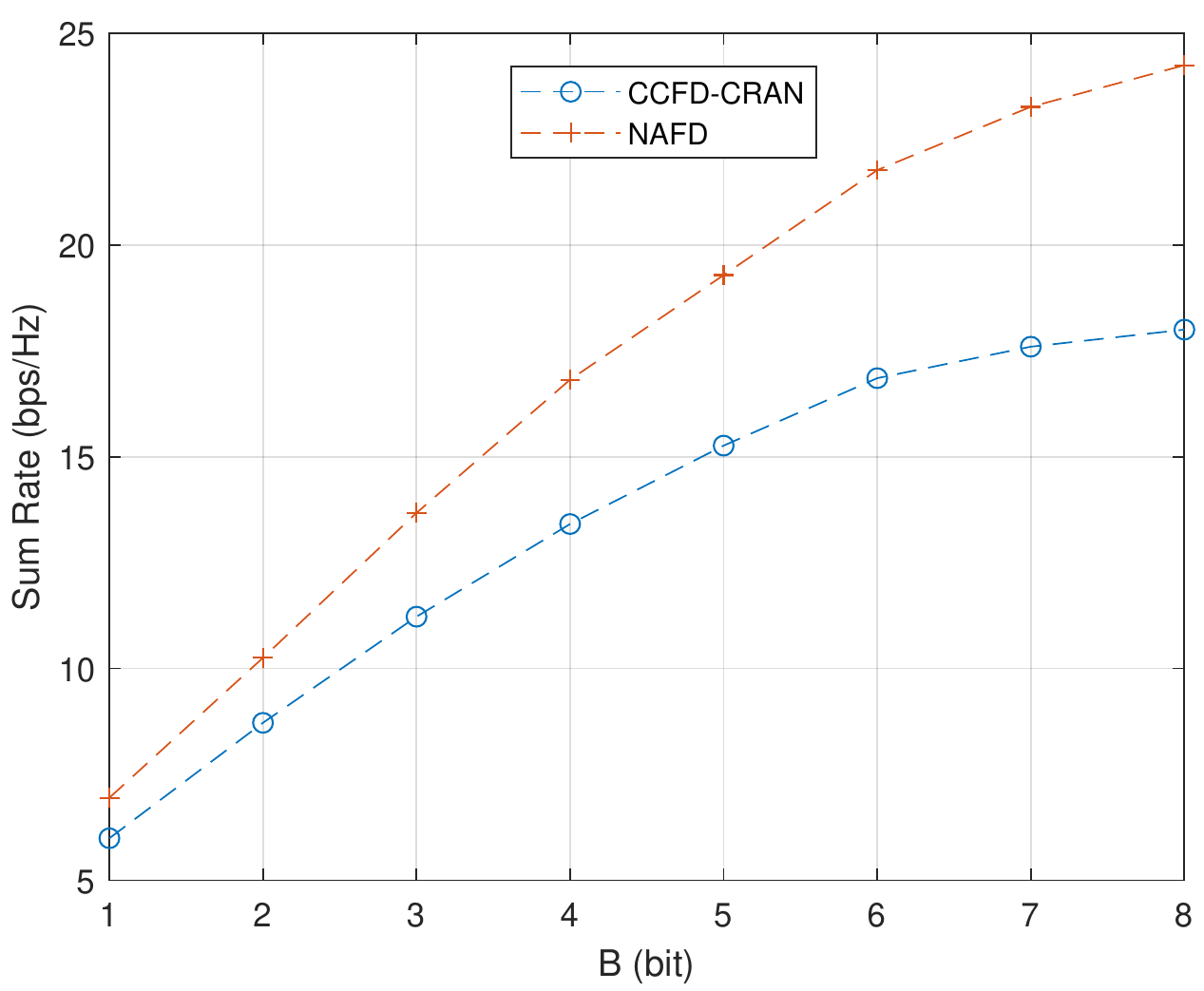}
		\caption{{\color{blue}Sum rate versus quantization bit $B$ with $C_D=C_U=130$ bps/Hz.}}
		\label{NAFD_CCFD}
	\end{figure}

	\begin{figure}
   \centering
   \subfloat[Comparison correspond to $1,...,4$ bit]{\includegraphics[width=3.3in]{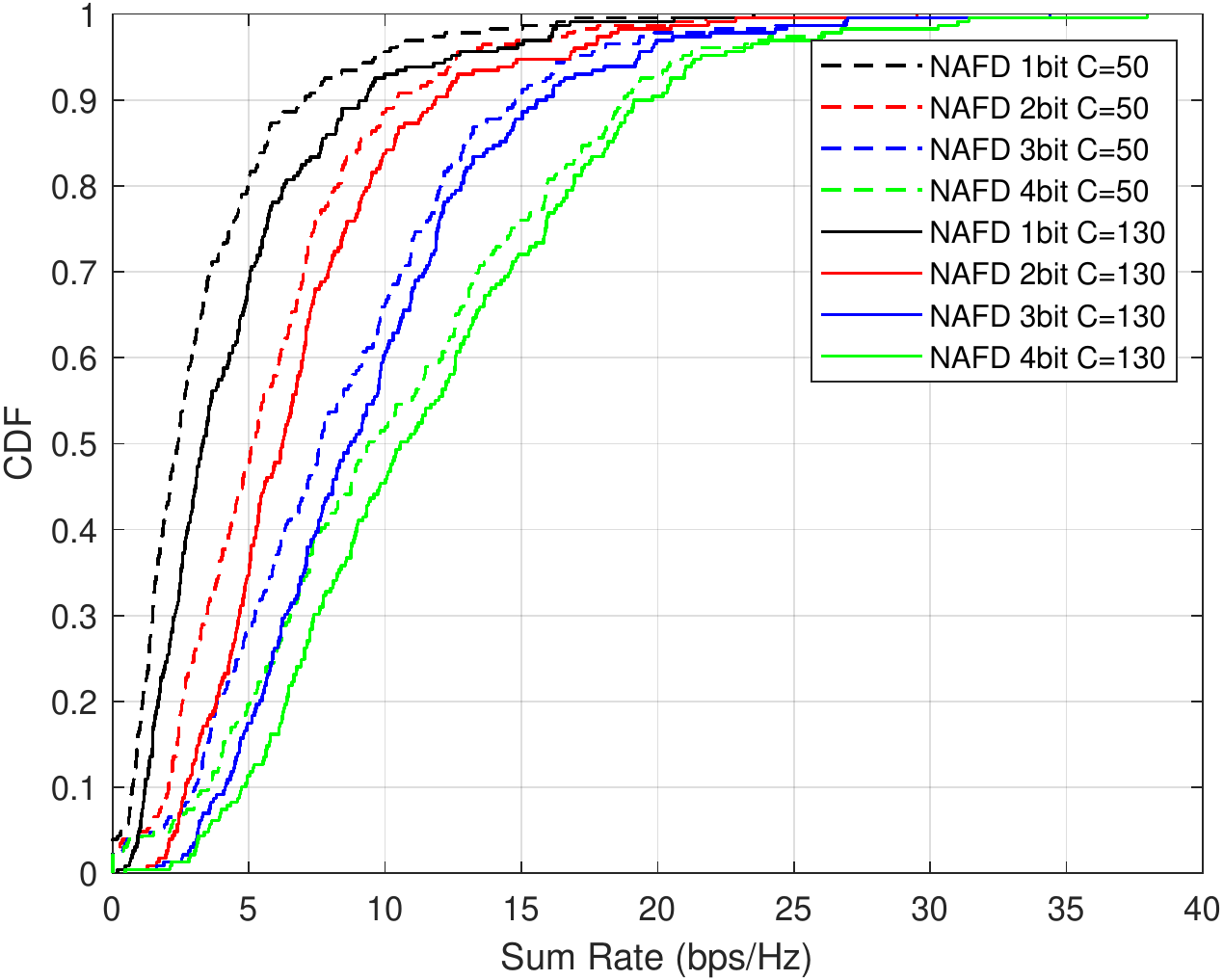}
   \label{1_4bit}} \\
   \subfloat[Comparison correspond to $5,...,8$ bit]{\includegraphics[width=3.3in]{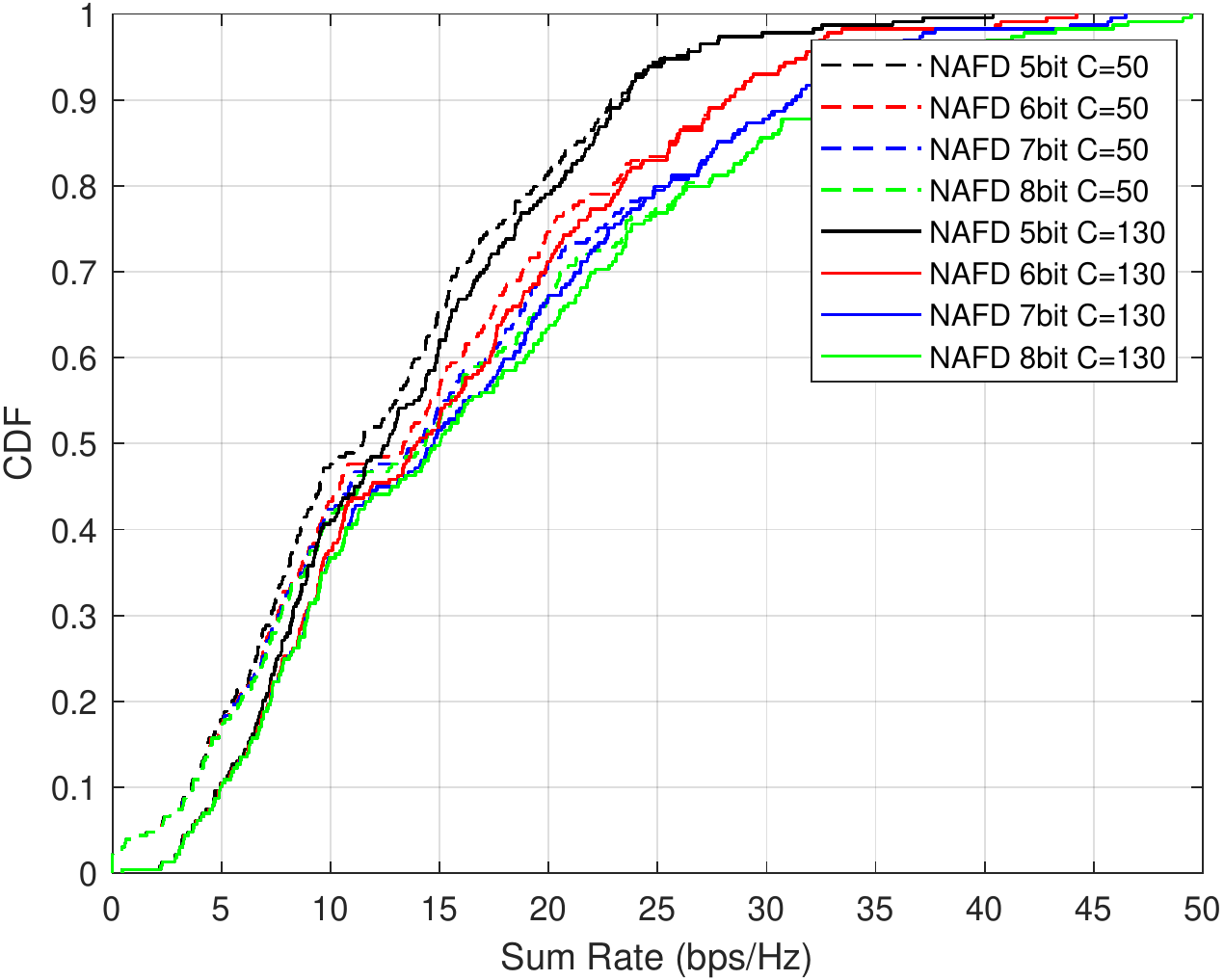}
   \label{5_8bit}}
   \caption{{\color{blue}Comparison of the cumulative distribution function (CDF) for the lower bounds of the achievable sum rate correspond to $B=1,...,8$ bit with $C_D=C_U=C=50$ and $C_D=C_U=C=130$ bps/Hz, respectively.}}
   \end{figure}
	
	Then, we compare NAFD cell-free mmWave massive MIMO system with DAC quantization and fronthaul compression to that with CCFD C-RAN, both of which are solved by the proposed algorithm. Fig. \ref{NAFD_CCFD} shows the sum rate in NAFD system and CCFD C-RAN system versus quantization bit $B$ with $C_D=C_U=130$ bps/Hz. It can be seen from Fig. \ref{NAFD_CCFD} that the sum rate in NAFD system and CCFD C-RAN system increases as quantization bit $B$ becomes higher. {\color{red}This is because increasing the quantization bit $B$ means a higher DAC resolution and a smaller quantization distortion factor, which results in less negative impact on the sum rate.} As far as the NAFD system is concerned, the sum rate is {\color{blue}$17.29$} bps/Hz higher when $B=8$ bit than that in the case of $B=1$ bit. {\color{blue}Furthermore, it also can be observed that the performance of the NAFD system is always better than that of the CCFD C-RAN system and their difference becomes larger as the quantization bit $B$ increases.} {\color{red}This is consistent with the theory that the performance comparison between NAFD and CCFD C-RAN is similar to the comparison between distributed MIMO and co-located MIMO. Since the RAUs are placed in more favorable positions in the NAFD system, distributed antennas can utilize spatial resources more efficiently and achieve additional power gains and macro diversity.} Specifically, the sum rate difference between NAFD and CCFD C-RAN system is {\color{blue}$0.96$} bps/Hz when quantization bit $B=1$ bit, and increases to {\color{blue}$6.24$} bps/Hz when $B=8$ bit, {\color{blue}which attains a 5.5-fold improvement.}

	Finally, Fig. \ref{1_4bit} and Fig. \ref{5_8bit} present the comparison of CDF for the lower bounds of the achievable weighted sum rate corresponding to $B=1,2,...,8$ bit with $C_D=C_U=C=50$ bps/Hz and $C_D=C_U=C=130$ bps/Hz, respectively. As observed from Fig. \ref{1_4bit} and Fig. \ref{5_8bit}, when the larger fronthaul capacity $C_D=C_U=C=130$ bps/Hz is at $B=1,...,4$ bit, the sum rate of the system are obviously greater than the smaller fronthaul capacity $C_D=C_U=C=50$ bps/Hz. In particular, the $C_D=C_U=C=130$ bps/Hz performs better than the $C_D=C_U=C=50$ bps/Hz in 100\% simulations, while both of them  are comparable at $B=5,...,8$ bit. This explains that the performance of the system at low quantization bit is more sensitive to fronthaul capacity than higher quantization bit.

	\section{Conclusion}
	In this paper, the NAFD cell-free mmWave massive MIMO system with DAC quantization and fronthaul compression is proposed. We put forward an iterative SCA-based algorithm to maximize the weighted uplink and downlink sum rate under the constraints of the bidirectional fronthaul compression capacity, and the power limitation of the T-RAUs and uplink users. The simulation results validate the convergence of the proposed algorithm and demonstrate that NAFD cell-free mmWave massive MIMO systems with DAC quantization and fronthaul compression perform better than that with CCFD C-RAN, especially in the case of high quantization bits.  We further study the impact of fronthaul capacity and DAC quantization on the spectral efficiency of the proposed NAFD system,  which draws a conclusion that increasing fronthaul capacity and DAC quantization bit contributes to realizing a higher achievable rate.

	
	%

	
		\bibliographystyle{IEEEtran}
		\bibliography{bare_jrnl}

	
	\ifCLASSOPTIONcaptionsoff
	\newpage
	\fi
	
\end{document}